\documentclass[11pt]{article}
\usepackage[]{moriond,epsfig}

\def\Journal#1#2#3#4{{#1} {\bf #2}, #3 (#4)}

\def\NIMA{{\em Nucl. Instrum. Methods} A}

\def\PLB{{\em Phys. Lett.}  B}

\def\be{\begin{equation}}
\def\ee{\end{equation}}
\def\bea{\begin{eqnarray}}
\def\eea{\end{eqnarray}}

\begin{document}
\vspace*{4cm}
\title{DETERMINATION OF THE POLARIZATION VECTOR OF POSITRONS FROM 
THE DECAY OF POLARIZED MUONS}

\author{ KAI-U. K\"OHLER for the ETH Z\"urich -- Cracow -- PSI -- 
Collaboration }

\address{ETH Z\"urich\,-\,Institute for Particle Physics, H\"onggerberg, 
CH-8093 Z\"urich , Switzerland}

\maketitle\abstracts{
In the standard model of electroweak interactions the positrons from 
the decay of polarized positive muons are mainly longitudinally polarized.  
The measurement of the two transverse polarization components 
$P_{\rm{T_{1}}}$, which lies in the plane spanned by  muon-spin and 
positron momentum, and $P_{\rm{T_{2}}}$, which is perpendicular to this
plane, is a sensitive tool to look for contributions from additional, 
exotic interactions. The energy dependence of $P_{\rm{T_{1}}}$ yields 
the low energy Michel parameter $\eta$ and thus an improved 
model-independent value of the Fermi coupling constant. A non-zero value 
of $P_{\rm{T_{2}}}$ would be the first observation of time reversal 
violation in a purely leptonic decay.
The muon decay experiment at the Paul Scherrer Institut is the 
first experiment to measure all three positron polarization components,
$P_{\rm{T_{1}}}$, $P_{\rm{T_{2}}}$, and the longitudinal polarization 
component $P_{\rm{L}}$. In this contribution results from 
a small data-sample taken in  1999 are presented. The values obtained 
are \mbox{$P_{\rm{T_{1}}} = (5 \pm 16 ) \times 10^{-3}$}, 
$P_{\rm{T_{2}}} = (1 \pm 16) \times 10^{-3}$, and $P_{\rm{L}} 
= 1.09 \pm 0.15$. More than ten times more data was taken in 
October\,/\,November 2000, the analysis of which will further reduce 
the statistical errors.}

\section{Introduction}

Though it has been shown fifteen years ago by Fetscher et 
al.~\cite{fet86} that $V-A$, which is one of the basic assumptions of 
the standard model, follows from the results of a selected set of muon 
decay experiments, the experimental limits obtained up to now still 
allow for substantial contributions from non-standard couplings. Limits 
on these couplings can be efficiently reduced by studying polarized 
muons and positrons.

The muon decay experiment~\cite{bar00} performed at PSI was designed for the 
measurement of the transverse polarization components of the positrons 
from the decay of polarized muons. Being able to additionally measure 
the longitudinal polarization, it is the first experiment to determine 
the complete polarization vector of the positrons.

One distinguishes two orthogonal transverse polarization components 
$P_{\rm T_{1}}$ and $P_{\rm T_{2}}$. $P_{\rm T_{1}}$ lies in a plane 
spanned by the muon polarization and the positron momentum, and 
$P_{\rm T_{2}}$ is perpendicular to that plane. Within the standard 
model the energy averaged value of $P_{\rm T_{1}}$ is $0.003$, whereas  
$P_{\rm T_{2}}$ is exactly zero. Both components can be expressed
in terms of the Michel parameters $\eta$, $\eta^{\prime \prime}$, 
$\frac{\alpha^{\prime}}{A}$ and $\frac{\beta^{\prime}}{A}$~\cite{fet95}.
For the~$V - A$~theory these four parameters are all zero.

However, assuming one additional coupling one can 
obtain substantial transverse polarization components of which 
$P_{\rm T_{2}}$ violates time reversal invariance. Neglecting exotic 
contributions in second order, one derives from the energy dependence 
of the two transverse polarization components
\be
	P_{\rm T_{1}}(E_{\rm e}) \rightarrow \eta \approx \frac{1}{2} \Re 
	\{g_{RR}^{S}\} \qquad \mbox{and} \qquad
	P_{\rm T_{2}}(E_{\rm e}) \rightarrow \frac{\beta^{\prime}}{A} 
	\approx \frac{1}{4} \Im \{g_{RR}^{S}\},
	\label{eq:gSRR}
\ee
where $g_{RR}^{S}$ represents a scalar, charge-changing 
interaction with right-handed charged leptons.

A fit to all currently available data~\cite{bur85} gives $\eta = (-7 
\pm 13) \times 10^{-3}$. A more precise value of $\eta$ is needed for 
a model-independent determination of the Fermi coupling constant 
$G_{\rm F}$\,: The influence of the uncertainty in the experimental 
value of $\eta$ on the value of $G_{\rm F}$ is at present $20$~times 
larger than the one of the muon lifetime~\cite{fet95}.

\section{The Experiment}
\subsection{Experimental Setup}
\begin{figure}[hb]
	\begin{center}
		\vspace{-6mm}
		\epsfig{file=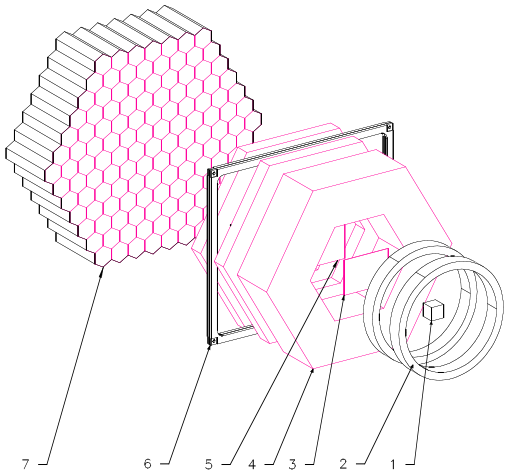, width=9.5cm}
		\vspace{-9mm}
	\end{center}
	\caption{Setup of the experiment. The various components are 
	described in the text.}
	\label{fig:exp}
	\vspace{-3mm}
\end{figure}

The experimental setup is shown in Fig.\ref{fig:exp}. Highly polarized 
muons arrive in bunches every $20$~ns and are stopped in a Be 
target~(1). Their polarization precesses in a homogeneous magnetic 
field induced by two Helmholtz coils~(2). A high polarization of the 
muons in the target is preserved by tuning the precession frequency to 
be equal to the accelerator RF. This ensures that the arriving muon 
bunches are added coherently to the muons already in the target. The 
decay $e^{+}$ passing the trigger scintillators~(3) and~(5) are 
tracked by a set of drift chambers~(4) and can annihilate with 
polarized $e^{-}$ in a magnetized foil~(6). The two annihilation 
quanta are detected as two separated clusters in a BGO 
calorimeter~(7) consisting of 127 hexagonal crystals. For more 
details on the apparatus, see ref.~(\,\cite{bar00}).

\subsection{Experimental Methods}
The polarization vector of the positrons can be split into components 
in the following way:
\be
    \vec{P}_{e^{+}} = \left(
     \begin{array}{c}
         P_{\mathrm{T}_{1}}  \\
         P_{\mathrm{T}_{2}}  \\
         P_{\mathrm{L}}
     \end{array}
     \right) 
 	\equiv 
     \left(
     \begin{array}{c}
 		P_{\rm T} \cdot \cos \varphi \\
 		P_{\rm T} \cdot \sin \varphi \\
 		P_{\mathrm{L}}	
      \end{array}	
    \right) 
\ee

The observables that define the complete vector are the absolute value 
$P_{\rm T}$ of the transverse polarization, a phase $\varphi$ and the 
longitudinal polarization component $P_{\rm L}$. In order to determine 
these three observables, three simultaneous but independent 
measurements are performed:

\subsection*{Measurement of $P_{\rm T}$ via the time dependence of 
annihilation}
The cross-section for annihilation between polarized $e^{+}$ and 
$e^{-}$ depends on the relative orientation of electron polarization 
and a possible transverse component of the positron polarization.
Due to the muon-spin rotation, a transverse positron polarization 
would be detected as a harmonic time dependence of the annihilation 
rate for a given BGO-detector pair which can be written in the form
\be
    f(t) = 1 + A \cdot \sin(\omega \,t + \alpha),
	\label{eq:timd}
\ee
where $A$ and $\alpha$ are functions of the energy, the position of 
the two annihilation quanta and two orthogonal transverse polarization components 
$P_{1}$ and $P_{2}$. Eq.~\ref{eq:timd} is used to perform a 
log-likelihood parameter estimation of $P_{1}$ and $P_{2}$.
As these components are selected regardless of
the orientation of the muon polarization, they only represent the 
absolute value of the transverse polarization.

\subsection*{Determination of the phase $\varphi$ of the transverse 
polarization using the $\mu$SR effect}
Since the accepted decay positrons are emitted into a cone with the 
axis being perpendicular to the precession plane of the muon 
polarization, there is a small remnant $\mu$SR effect (i.e. a 
time-dependent rate variation due to the decay asymmetry with respect 
to the precessing muon spin~\cite{fet95}). One uses this effect to 
determine time zero, i.e. the position of the precessing polarization 
vector $\vec{P}_{\mu}$ of the muon. Since $P_{\rm T_{1}}$ and 
$P_{\rm T_{2}}$ are defined relative to $\vec{P}_{\mu}$, this finally 
allows to determine the two components separately.

\subsection*{Measurement of the longitudinal polarization $P_{\rm L}$ 
via the spatial dependence of annihilation}
Positrons hitting the magnetized foil off the symmetry axis have a 
component of their longitudinal polarization in the direction of the 
electron polarization. The annihilation cross-section depends on the 
relative orientation of these two components, which can be either 
parallel or antiparallel. Thus, by dividing the fiducial area of the 
magnetized foil into rectangular bins and calculating the asymmetry 
of the annihilation rate for each bin resulting from reversing the 
foil magnetization, one can deduce the longitudinal polarization. 

\section{Preliminary Results}
In a first data taking run in fall of 1999, $240 \times 10^{6}$ annihilation-like 
events were recorded. Appropriate geometry and energy cuts efficiently 
reduce the background and yield a data sample of $11 \times 10^{6}$ 
valid annihilation events on which the following results are based.

\begin{figure}[ht]
	\begin{center}
		\epsfig{file=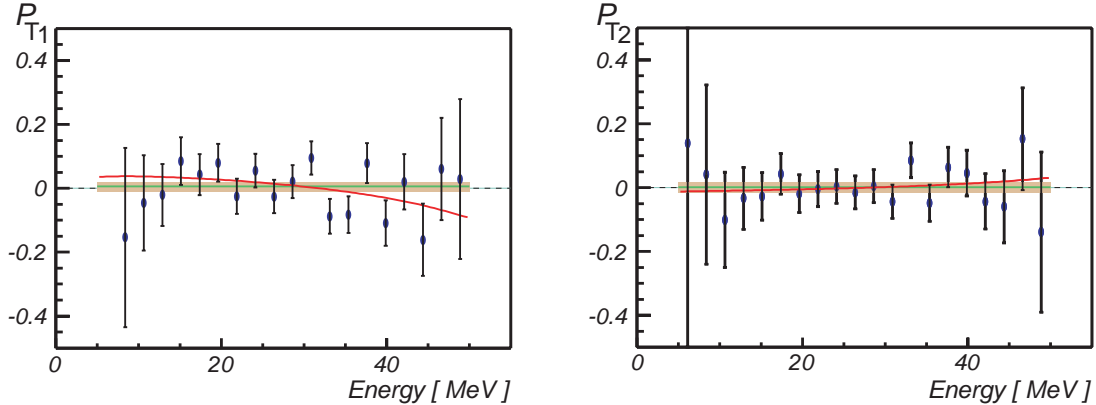, width=14.5cm}
 		\vspace{-2mm}
	\end{center}
	\caption{Transverse polarization components $P_{\rm T_{1}}$ and 
	$P_{\rm T_{2}}$ at the time of annihilation. The horizontal lines 
	indicate the averages, the boxes correspond to their errors. The 
	curved lines represent the results of fits to the energy dependence 
	of $P_{\rm T_{1}}$ and $P_{\rm T_{2}}$ to determine the Michel 
	parameters.\hspace*{5.7cm}}
	\label{fig:pt}
\end{figure}

The transverse polarization components as a function of energy are 
shown in Fig.~\ref{fig:pt}. The energy-averaged values for the two 
transverse polarization components are 
$<P_{\rm T_{1}}> = 0.005 \pm 0.016_{\rm stat} \pm 0.003_{\rm sys}$ and 
$<P_{\rm T_{2}}> = 0.001 \pm 0.016_{\rm stat} \pm 0.003_{\rm sys}$.
The statistical errors are by a factor of 1.3 smaller than the current 
experimental limits and corroborates the findings of~(\cite{bur85}).

The result for the longitudinal polarization is $P_{\rm L} = 1.09 \pm 
0.15$. This agrees well with the prediction $P_{\rm L} = 1$ of the 
standard model and confirms that our apparatus is sensitive to 
positron polarization.

Using the measured energy dependence of $P_{\rm T_{1}}$ and 
$P_{\rm T_{2}}$, the functions describing the transverse polarization 
components in terms of the Michel parameters can be fitted (see 
Fig.~\ref{fig:pt}). As it is known~\cite{fet86} that the weak 
interaction is dominated by $V - A$, it 
is explicitly assumed, that only one additional coupling contributes 
to the transverse positron polarization, which then has to be $g_{RR}^{S}$.
This leaves only $\eta$ and $\frac{\beta^{\prime}}{A}$ as free 
parameters for the fit. This assumption is supported by small values of 
$\alpha$ and $\alpha^{\prime}$~\cite{bur85b}, which implies $\eta 
\approx - \eta^{\prime \prime}$. The results of this fit are $\eta = 
-0.004 \pm 0.014_{\rm stat} \pm 0.002_{\rm sys}$ and 
$\frac{\beta^{\prime}}{A} = 0.001 \pm 0.007_{\rm stat} \pm 0.001_{\rm sys}$.

\section{Conclusion and Outlook}
From the analysis of $11 \times 10^{6}$ annihilation events one 
obtains new limits for $P_{\rm T_{1}}$ and $P_{\rm T_{2}}$. Based on 
the Michel parameters $\eta$ and $\frac{\beta^{\prime}}{A}$ resulting 
from a fit to the energy dependence of these two positron polarization 
components, limits on the coupling constant $g_{RR}^{S}$ for a 
scalar, charge-changing interaction with right-handed leptons can be 
derived. According to eq.~(\ref{eq:gSRR}) one obtains 
$\Re \{g_{RR}^{S}\} = -0.008 \pm 0.028$ and 
$\Im \{g_{RR}^{S}\} = 0.004 \pm 0.028$, respectively 
$|g_{RR}^{S}| = 0.009 \pm 0.028$.

All results presented agree with the expectations from the standard 
model.
However, in a longer data taking run in October/November 2000 the 
event statistics were improved by more than a factor of 10. The 
analysis of this data will improve all quoted limits or give a hint 
for physics beyond the standard model.

\section*{Acknowledgments}
I would like to thank all the members of the collaboration between ETH 
Zurich, PSI and the three institutes in Poland. Detailed information 
about the participating scientists and institutions can be found on 
the web-page {\em http://pc2107.psi.ch/collaboration.html}.

This project is supported in part by the Swiss National Science Foundation
and by the Polish Committee for Scientific Research under Grant No. 
2P03B05111.

\end{document}